\documentclass[aps,pra, superscriptaddress, reprint, longbibliography]{revtex4-1}

\usepackage{graphicx}
\usepackage{soul}
\usepackage[colorlinks,
			linkcolor=blue,
			urlcolor=blue,
			citecolor=blue]{hyperref}
\usepackage[T1]{fontenc}       % Use modern font encodings
\usepackage[version=3]{mhchem}
\usepackage{enumerate}
\usepackage{titlesec}
\titlespacing*{\section}{0pt}{1.1\baselineskip}{\baselineskip}

\begin{document}
		
	\title[]{Influence of thermal boundary conditions on the\\current-driven resistive transition in \ce{VO2} microbridges}
	
	\author{Nicola \surname{Manca}}
	\email{n.manca@tudelft.nl, nicola.manca@spin.cnr.it}
	\affiliation{\mbox{Physics Department, University of Genova, Via Dodecaneso 33, Genova 16146, Italy}}
	\affiliation{\mbox{CNR-SPIN, Corso Perrone 24, Genova 16152, Italy}}
	\affiliation{\mbox{Kavli Institute of Nanoscience, Delft University of Technology, P.O. Box 5046, 2600 GA Delft, The Netherlands}}
	
	\author{Teruo \surname{Kanki}}
	\affiliation{\mbox{Institute of Scientific and Industrial Research, Osaka University, Ibaraki, Osaka 567-0047, Japan}}
	
	\author{Hidekazu \surname{Tanaka}}
	\affiliation{\mbox{Institute of Scientific and Industrial Research, Osaka University, Ibaraki, Osaka 567-0047, Japan}}
	
	\author{Daniele \surname{Marr\'e}}
	\affiliation{\mbox{Physics Department, University of Genova, Via Dodecaneso 33, Genova 16146, Italy}}
	
	\author{Luca \surname{Pellegrino}}
	\affiliation{\mbox{CNR-SPIN, Corso Perrone 24, Genova 16152, Italy}}
	
	\begin{abstract}
		We investigate the resistive switching behaviour of \ce{VO2} microbridges under current bias as a function of temperature and thermal coupling with the heat bath. Upon increasing the electrical current bias, the formation of the metallic phase can progress smoothly or through sharp jumps. The magnitude and threshold current values of these sharp resistance drops show random behaviour and are dramatically influenced by thermal dissipation conditions. Our results also evidence how the propagation of the metallic phase induced by electrical current in \ce{VO2}, and thus the shape of the resulting high-conductivity path, are not predictable. We discuss the origin of the switching events through a simple electro-thermal model based on the domain structure of \ce{VO2} films that can be useful to improve the stability and controllability of future \ce{VO2}-based devices.\\
		
		\noindent\textcopyright~2015 AIP Publishing LLC.\\
		Published online 8 October 2015 (\url{http://dx.doi.org/10.1063/1.4933014})
		
	\end{abstract}
	
	\maketitle
	
	Vanadium Dioxide (\ce{VO2}) is particularly suitable for the development of tuneable electronic devices because of its hysteretic phase transition (PT) occurring above room temperature (approximately 65\,$^\circ$C). This PT affects crystal symmetry, electrical resistivity (lowered by about four orders of magnitude), values of the optical constants, thermal conduction and specific heat \cite{Zylbersztejn1975, Leroux1998, Eyert2002, Qazilbash2009, Tomczak2009, Oh2010}. The origin of the PT has been addressed to both structural and electronic correlation effects and is still under debate \cite{Wentzcovitch1994, Rice1994, Biermann2005, Kumar2014, Budai2014a, Morrison2014}. From the applicative point of view, its unique characteristics led to the development of several prototypical systems, such as memristors \cite{Lee2007, Driscoll2009a, Driscoll2010}, micro-actuators \cite{Tselev2011, Merced2015}  or optical devices \cite{Coy2010, Kats2012, Kats2014a}. Manipulation of \ce{VO2} domains at the micro- and nanoscale through an electrical bias has been recently demonstrated \cite{Kanki2012}. Nevertheless, domains manipulation by electric biases can be associated to an increase of temperature due to Joule effect or to carrier injection \cite{Joushaghani2014, Rathi2014}. In the framework of developing \ce{VO2} devices driven by electrical bias, it has been shown that current biasing increases \ce{VO2} operation lifetime \cite{Crunteanu2010} and allows the fine control of the metallic/insulating domains ratio on micrometric regions, enabling the stable and reproducible addressing of multiple resistive and mechanical states \cite{Merced2015, Pellegrino2012, Manca2013}. A systematic analysis of the transition dynamics in \ce{VO2} micro-devices under current bias is lacking so far, while both sharp \cite{Kim2012, Jordan2014, Zhang2014a, Aliev2014b} and smooth transition \cite{Pellegrino2012, Manca2013, Zhong2011, Kumar2013} have been experimentally observed.
	
\begin{figure}
		\includegraphics[width=1.0\linewidth]{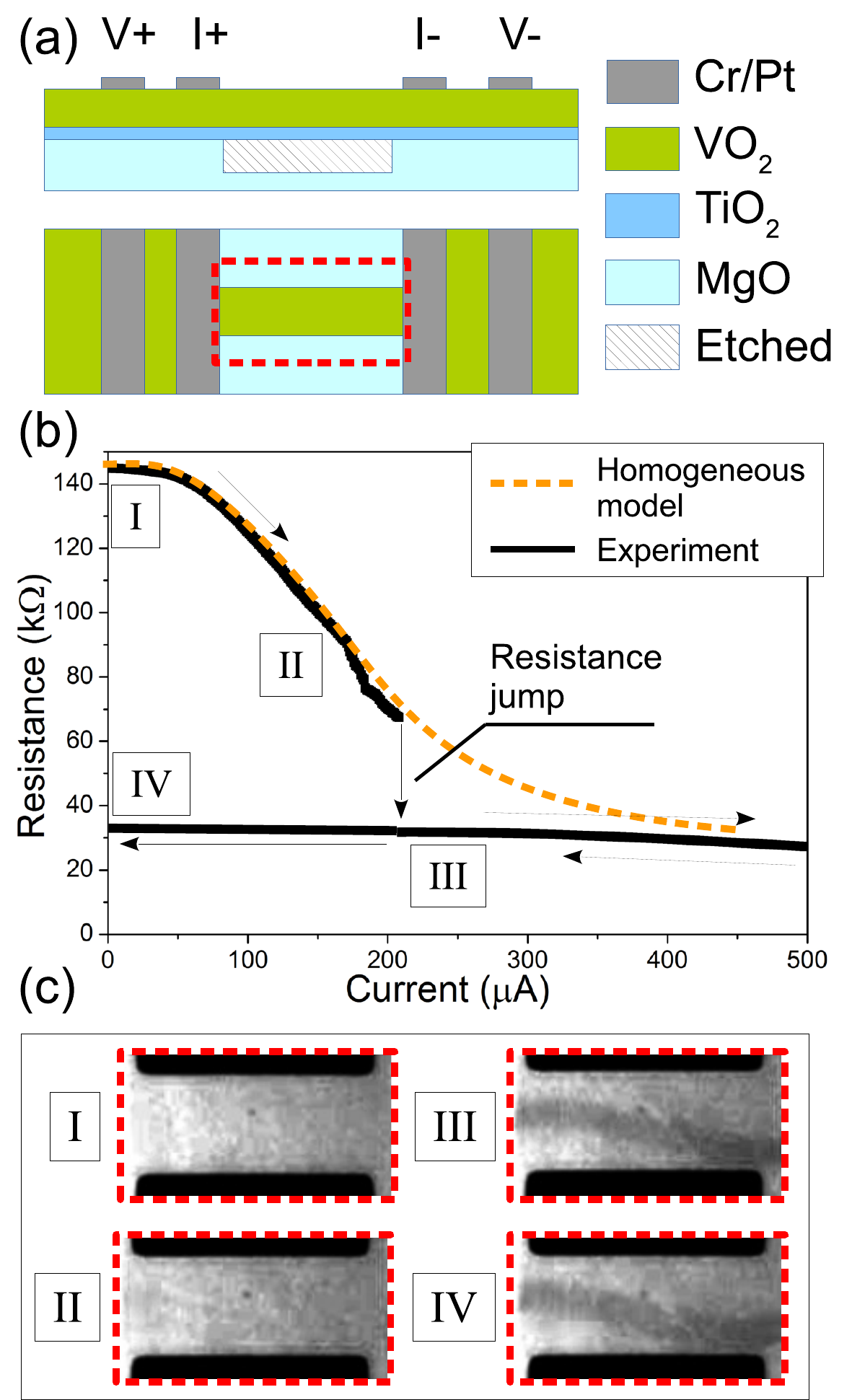}
		\caption{\label{fig:Fig1}
		(a) Sketch of a \ce{VO2}/TiO$_2$ microbridge, side and top view. Channel dimensions are $\mathrm{8\, \mu m\, \times\, 20\, \mu m\, \times\, 70\, \mu m}$. The dashed rectangle corresponds to the optical images reported in (c). In the free-standing case, the shaded region is removed by selective wet etching.
		(b) R(I) plot for a clamped microbridge compared to what expected in a homogeneous-medium approximation.
		(c) Optical images of the clamped microbridge during the current ramp reported in (b) (See Supplementary Material). The metallic strip is still present at zero current (IV), because of the hysteresis of \ce{VO2}. Thermostat temperature is set to 66\,$^\circ$C.}
	\end{figure}

	In this letter, we investigate the resistive switching behaviour of \ce{VO2} micro-bridges (MBs) under current bias. Our findings show that both sharp and smooth transitions are possible and that the current-induced PT in \ce{VO2} thin films is characterized by intrinsic random behaviour, without a predictable shape of the metallic \ce{VO2} channel upon switching and a reproducible value of the associated threshold current. We also show that switching characteristics are strongly influenced by the thermal environment and different behaviours can be observed by varying temperature or thermal coupling conditions. In order to systematically study this phenomenon, we performed a series of current ramps on a MB at different temperature setpoints. Thermal couplings conditions were modified by making the same MB free-standing on a second stage, by removing the shadowed portion of the MgO substrate of Fig.\,\ref{fig:Fig1}a. Out-of-plane thermal dissipation was also varied by changing the environmental pressure. Three configurations have been thus explored in sequence: clamped MB in air/free-standing MB in air/free-standing MB in vacuum (10$^{-2}$ Pa).
	
	\ce{VO2} MBs ($\mathrm{8\,\mu m \times 20\,\mu m}$) were fabricated by standard optical lithography from a \ce{VO2}(70\,nm)/\ce{TiO2(110)}(20\,nm) heterostructure grown on an \ce{MgO(001)} substrate by Pulsed Laser Deposition, similarly to what reported in Ref. \onlinecite{Pellegrino2012, Manca2013, Yamasaki2014}. \ce{TiO2} grown on top of MgO(001) is made of domains having two possible planar orientations  and typical dimension of tens of nanometers \cite{Okimura2005}. Because of the epitaxial growth of \ce{VO2} on \ce{TiO2} this is also the expected typical size of \ce{VO2} domains. Cr/Pt electrodes were used to obtain ohmic contacts and fabricated by sputtering and lift-off method. A schematic representation of the device and its electrical connections are presented in Fig.\,\ref{fig:Fig1}a: four-probes measurements were carried on by applying the electrical current next to the micro-bridge and by measuring the voltage at the V(+)(--) contacts. This configuration lowers the total resistance between the I(+)(--) contacts, decreasing the voltage required at high biasing, that could results in undesired electrochemical reactions. Free-standing structures were realized by soaking the MB in a \ce{H3PO4}(8.5\%) acid bath and drying using a critical point dryer and no degradation in \ce{VO2} film quality was observed (see Supplementary Material). Measurements were performed by four probe method in a vacuum chamber with controlled atmosphere and temperature, as described elsewhere \cite{Manca2013}. Because of the hysteretic nature of \ce{VO2}, the PID parameters of the heater were carefully tuned in order to avoid any possible overshoot during the heating of the devices.  Fig.\,\ref{fig:Fig1}b reports a typical Resistance vs Current plot for the clamped case. This measurement was performed at a fixed thermostat temperature of 66\,$^\circ$C, within the hysteresis loop of the material. The current bias was slowly increased at a rate of $\mathrm{1\,\mu A/0.75\,s}$, while the voltage drop was continuously monitored. The electrical resistance R(I) smoothly decreases up to about 200\,$\mu$A, then a sharp transition is detected and no further remarkable variation is observed up to 500\,$\mu$A. When decreasing the current to zero, the measured resistance remains almost constant, as expected from the hysteretic nature of \ce{VO2} at this temperature. We notice that the observed sharp transition of the electrical resistance cannot be explained by considering \ce{VO2} as a homogenous medium with negative temperature coefficient of resistance. In this case only smooth transitions are expected, as represented by the sketched dashed line of Fig.\,\ref{fig:Fig1}b. On the contrary, thermal runaway is present under voltage biasing, as it has been widely reported in literature for the cases of voltage-induced resistive switching of \ce{VO2} devices \cite{Joushaghani2014, Rathi2014, Kumar2013, Cao2010a, Kim2010, Stabile2014}. We acquired a series of optical images of our microbridges during the current biasing to highlight the relationship between the progressive formation of the metallic phase and the variations of the electrical resistance. The optical images of the microbridge in Fig.\,\ref{fig:Fig1}c progress as follows:
	\setlength{\leftmargini}{1.5em}
	\begin{enumerate}[I]
		\setlength{\itemindent}{-0.5em}
		\item The MB is initially in a full insulating state (light grey colour);
		\item Upon increasing current, the surface becomes slightly darker because of the random formation of metallic clusters;
		\item In conjunction with the resistance drop, we observe the formation of a dark metallic path connecting the two Cr/Pt electrodes;
		\item The metallic path is maintained even at zero current bias, in agreement with the electrical measurements.
	\end{enumerate}

	\begin{figure}
	\includegraphics[width=1.0\linewidth]{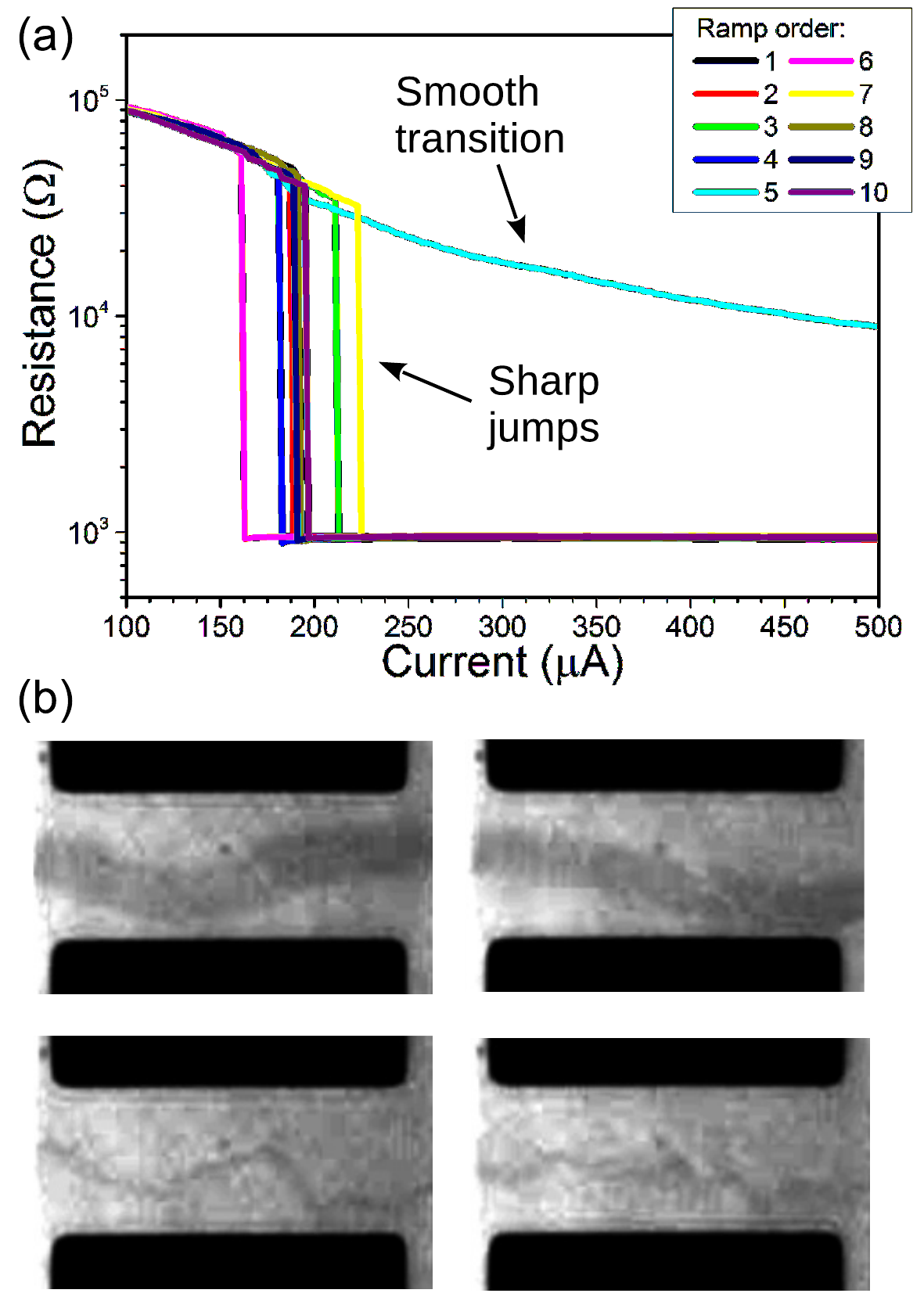}
	\caption{\label{fig:Fig2}
		(a) Series of R(I) plots measured on a clamped microbridge ($\mathrm{I_{bias}=0\,\mu A - 500\,\mu A}$) showing sharp and smooth transitions.
		(b) Optical images of metallic channels formed during different current ramps on the same clamped device. The conductive paths show random width and shape. Each current ramp is acquired after a cooling cycle down to room temperature which resets the initial state. Thermostat temperature is 66.5\,$^\circ$C in (a) and 66\,$^\circ$C in (b).}
	\end{figure}

\begin{figure*}[t]
	\begin{minipage}[c]{\textwidth}
		{
			\begin{minipage}[t]{0.70\textwidth}
				\mbox{}\\[-\baselineskip]
				\includegraphics[width=1.0\textwidth]{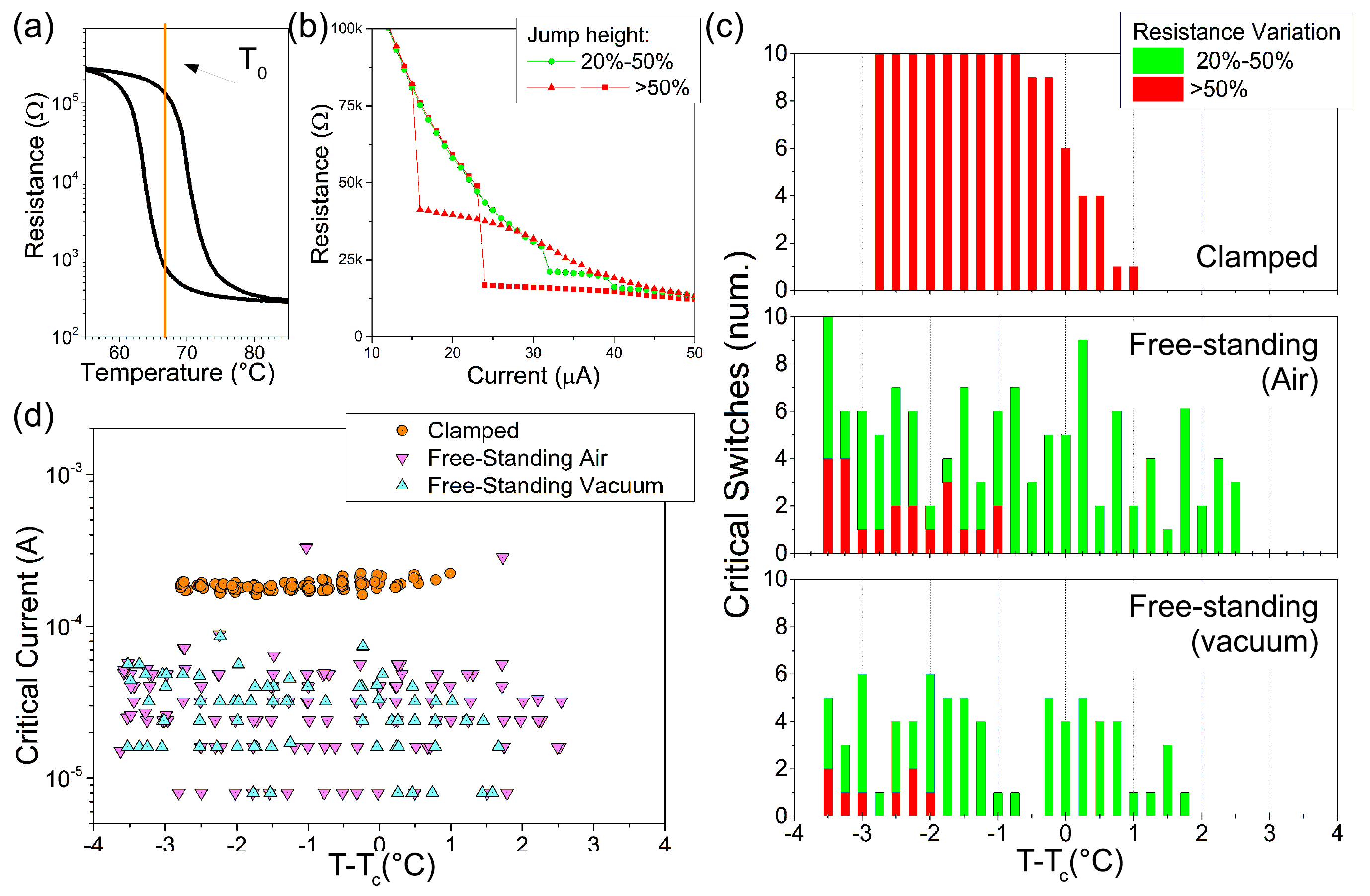}
			\end{minipage}\hfill
			\begin{minipage}[t]{0.25\textwidth}
				\mbox{}\\[-\baselineskip]
				\caption{\label{fig:Fig3}
					R(T) plot for the clamped \ce{VO2} microbridge. The orange line and the yellow shade indicate the center of the hysteresis (T$_0$=66.75\,$^\circ$C) and the explored temperature span of $\pm 4\,^\circ$C, respectively.
					(b) Examples of R(I) measured on the free-standing MB in air (T$_0$=68.5\,$^\circ$C).
					(c) Histograms of current ramps exhibiting at least one resistance jump, out of ten, for the same device at different temperatures and with different thermal coupling conditions. Two relative resistance variation values have been considered: above 20\% (green bars) and above 50\% (red bars).
					(d) Threshold current values at the resistive jumps of (c).}
			\end{minipage}
		}
	\end{minipage}
	\end{figure*}

	The formation of the metallic paths and the corresponding resistance drop apparently behave like intrinsically random phenomena. Fig.\,\ref{fig:Fig2}a reports a series of R(I) curves observed during different current ramps ($\mathrm{0\,\mu A - 500\,\mu A,\,1\,\mu A/0.75\,s}$). The thermal bath is at 66.5\,$^\circ$C and the system is cooled down to room temperature after each ramp. The first sharp resistive transition occurs at a ``critical current'' (I$\mathrm{_c}$) which varies by approximately 30\% across the different ramps and only in one case we observed a smooth transition. Fig.\,\ref{fig:Fig2}b reports optical images of different metallic channels on the clamped \ce{VO2} micro-bridge acquired after the transition. These channels present a variety of widths and shapes, even if they formed under the same experimental conditions (temperature and ramp speed). Notably, we observed no correlation between channel width and current threshold values. We also note that all the metallic channels show similar resistance, suggesting that the density of the metallic domains within the channels may be variable. The presence of these metallic channels clearly indicates how only a fraction of the \ce{VO2} domains is switching to the metallic phase, consistently with the lower resistance variation observed during the current-induced transition (Fig.\,\ref{fig:Fig2}a) with respect to that measured during the temperature sweep (Fig.\,\ref{fig:Fig3}a and Supplementary Material) and it is also in agreement with previous works \cite{Ha2013, Markov2015}. A typical modelling for \ce{VO2} considers the smooth R(T) observed in thin films as the results of the average contribution of a set of domains having single-crystal behaviour (sharp R(T)), with random distribution of their critical temperature or voltage \cite{Driscoll2012}. The switching condition of these single domains should be mainly determined by the elastic coupling with their neighbours and the substrate or by local defects that can efficiently shift the transition temperature \cite{Park2013}. In this framework, the transition should start from the specific region where the threshold conditions are lower, propagating accordingly with the distribution of the critical temperature among the different domains. The observed random formation of the metallic path indicates that, at least in our samples, the local fluctuations of temperature dominate over the distribution of the critical temperature values of the domains constituting the MB.
		
	We studied the switching behaviour of the MBs in three dissipative configurations. Measurements have been performed by using the following sequence:
	\begin{enumerate}
		\setlength{\itemindent}{-0.3em}
		\item Starting below the transition temperature (30\,$^\circ$C), the system is heated up to the temperature set-point (T$\mathrm{_{SET}}$) with a 120\,s waiting time for full thermalization;
		\item The current is increased from 0 $\mu$A to 500\,$\mu$A with steps of 1\,$\mu$A/0.75\,s;
		\item The temperature is lowered down to 30\,$^\circ$C.
	\end{enumerate}
	
	This sequence was repeated ten times for each T$\mathrm{_{SET}}$, then T$\mathrm{_{SET}}$ was increased by 0.25\,$^\circ$C and the sequence started over. We investigated a temperature span of about 4\,$^\circ$C above and below T$_0$, which is defined as the middle point between the maximum slopes of the logarithmic derivative of R(T) (see Fig.\,\ref{fig:Fig3}a) in the heating and cooling branches. We note that T$_0$ is 66.75\,$^\circ$C when the MB is measured in the clamped case, while it increases to 68.5\,$^\circ$C for the free-standing cases because of the weaker coupling with the thermal bath, which also slightly affects the R(T) shape \cite{Pellegrino2012, Manca2013, Yamasaki2014} (see Supplementary Material). From each sequence we extracted the following parameters: the critical current I$\mathrm{_c}$, the magnitude of the resistive variation when the sharp transition occurs ($\mathrm{\Delta R/R}$) and the number of ramps that present at least one significant sharp transition (jump) out of a total of ten (see after). Figure\,\ref{fig:Fig3}b shows three different representative current ramps obtained on a free-standing MB in air at the same temperature. Results from all the R(I) ramps, collected at each temperature and for the three configurations, are presented in Fig.\,\ref{fig:Fig3}c and d. In the clamped case the switching behaviour is very reproducible: all the jumps determine a dramatic lowering of the resistance similarly to what is observed in Fig.\,\ref{fig:Fig2}a and all the I$\mathrm{_c}$ lie around 200\,$\mu A$. Well below T$_0$, sharp transitions are always present, fading nearby T$_0$ toward R(I) plots characterized by a smooth lowering of the resistance. Free-standing cases present more complex behaviour and no clear transition from a sharp to smooth transitions regime with increasing T$\mathrm{_{SET}}$ is observed. We typically found that, at each T$\mathrm{_{SET}}$ value, different alternative behaviours can be observed:
	\begin{itemize}%[leftmargin=0pt]
		\setlength{\itemindent}{-0.5em}
		\item No transition at all, with smooth decrease of resistance; 
		\item Presence of several jumps having low magnitude;
		\item Sharp transitions similar to the clamped case.		
	\end{itemize}

	We stress that these three cases can be observed on the same device at the same thermal bath temperature (T$\mathrm{_{SET}}$). We also remind that after each current ramp we restored the low temperature insulating state of the \ce{VO2} film by lowering the thermostat temperature down to 30\,$^\circ$C. We counted/discarded the presence of a resistive jump event in the R(I) curves considering the magnitude of the relative resistance variation. In Fig.\,\ref{fig:Fig3}c we classify the switching events by using two thresholds values: $\mathrm{\Delta R /R > 20\%}$ and $\mathrm{\Delta R /R > 50\%}$. Lower jumps ($\mathrm{\Delta R /R < 20\%}$) have not been taken into account. The R(I) plots reported in Fig.\,\ref{fig:Fig3}b are coloured in order to identify the relative subset. Data points of Fig.\,\ref{fig:Fig3}d show the I$\mathrm{_c}$ value of the first jump having $\mathrm{\Delta R /R > 20\%}$ for each current ramp of the ensemble at a given T$\mathrm{_{SET}}$. In the free-standing case the switching behaviour is much more varied and sharp transitions can be observed in a broader temperature range of the hysteresis window. However, they are typically of lower magnitude and high-magnitude ones disappear at a temperature lower than that of the clamped case. We also note that often no sharp transition is observed for the entire R(I) curve, even at relative low temperature, where about 40\% of the ramps present smooth behaviour. The reduction of the number and magnitude of resistance jump events is accompanied by an increase of the temperature span where they can be still detected. In fact, we measured resistance jumps of small magnitude even at relative high temperature, where just smooth transitions can be observed for the clamped case.  As a general trend, if out-of-plane thermal dissipation is lowered by making the structure free-standing and reducing the gas pressure, sharp transitions decrease both in number and amplitude, as evidenced by the histograms of Fig.\,\ref{fig:Fig3}c. Our data also indicate that free-standing cases have a lower predictability of the jumps magnitude and critical current value. We attribute the differences between the three configurations to thermal effects. The presence of jumps depends on the specific experimental conditions such as working temperature, current bias and device geometry. For instance, no significant switching is reported in our previous works on \ce{VO2}-based devices, where the presence of smooth transitions allowed to modify with continuity the state of the devices \cite{Pellegrino2012, Manca2013}.  
	
\begin{figure}[t]
		\includegraphics[width=1.0\linewidth]{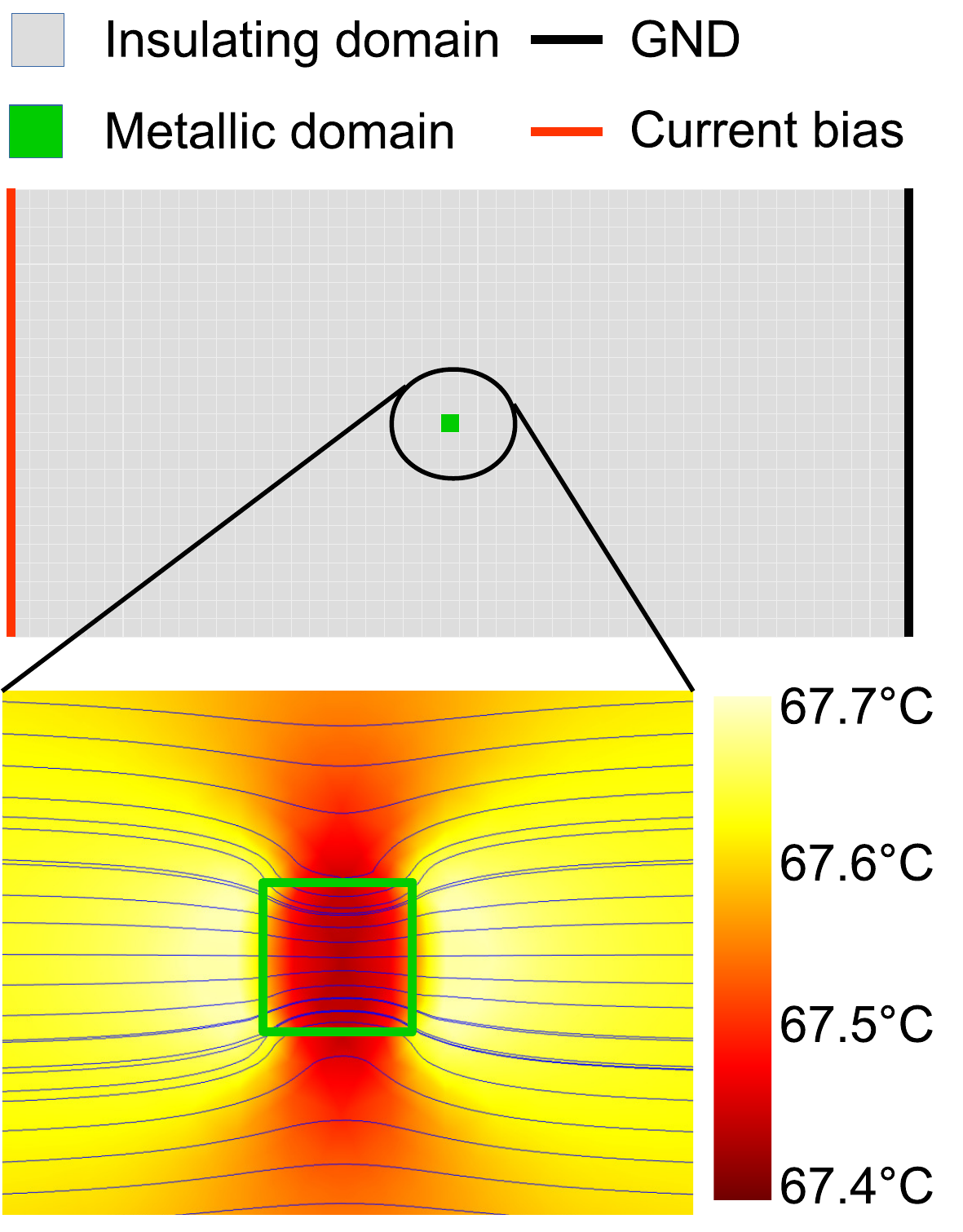}
		\caption{\label{fig:Fig4}
			Finite elements analysis showing the pinch effect on local current density from a metallic \ce{VO2} domain (black square) surrounded by insulating domains $\mathrm{\sigma_{ins}=10^2\,S/m,\ \sigma_{met}=10^5\,S/m}$.}
	\end{figure}

	As previously discussed, a homogeneous-medium thermal model is unable to reproduce the observed resistance switching. It  is thus required to consider the multi-domain structure of \ce{VO2} films. The initial smooth resistance lowering, observed during the current ramps (Fig.\,\ref{fig:Fig1}b), is compatible with the formation of randomly-arranged \ce{VO2} domains driven to the metallic phase by Joule heating. The switching events can be instead explained as an avalanche phenomenon triggered by the strong pinch effect \cite{Volkov1969, Wacker1995, Alekseev1998} of these metallic domains on the local current density distribution. This is shown in Fig.\,\ref{fig:Fig4}, where finite elements analysis (Comsol\textregistered~4.3b) of the temperature distribution and electrical current flow for a metallic domain ($\mathrm{\sigma_{met}=10^5\ S/m}$) surrounded by an insulating matrix ($\mathrm{\sigma_{ins}=10^2\ S/m}$) in the clamped case is reported. The clamping condition has been modelled by setting on the back of the 100\,nm thick domains an heat exchange coefficient of h=10000\,W/(m$^2$K) toward a 340\,K heat bath.
	
	Because of redistribution of the current, in-/off-axis domains experience a temperature increase/lowering with a relative difference of about 0.2\,K and an almost doubled current streamlines density. This value is strongly influenced by the simulation parameters, but simulation shows how few disconnected metallic domains, giving almost no macroscopic effect, can effectively modify their local thermo-electric environment triggering a catastrophic avalanche effect. This mechanism may act synergically with the fast switching phenomena occurring by Poole-Frenkel-assisted heating observed for vanadium oxides \cite{Rathi2014, Markov2015, Yang2011b, Brockman2014}. Our model however suggests that even small temperature fluctuations can initiate the avalanche process if the device is nearby the IMT temperature, since pinch effect determines an increase of local temperature for the in-axis domains and the metallic phase propagates until the two Cr/Pt electrodes are connected (Fig.\,\ref{fig:Fig2}b) with consequent sharp resistance drop. Because of the electro-thermal nature of this effect, the dissipative conditions are fundamental in determining switching dynamics. In the clamped case, the thermal resistance at any point of the MB is almost independent on the position, being every domain in direct contact with the substrate that acts as heat sink. Instead, in free-standing structures each domain has a different thermal resistance given by the distance from the heat sinks at the edges.  The temperature at the centre of the MB is easily increased by a small amount of current and its distribution along the structure is strongly inhomogeneous because of the different magnitude of the in-plane and out-of-plane dissipations \cite{Ceriale2014}. We believe that this temperature gradient is the limiting factor in the propagation of the metallic phase, which ends up in a series of small partial jumps instead of a sudden and wide change as in the clamped case.
	
	We finally note how the random shape of the metallic path in Fig.\,\ref{fig:Fig2}b indicates that multiple possible evolutions of the system are possible when the avalanche process starts. This is allowed if the differences among the domains are small, otherwise the path shape would be always identical and determined by the position of the domain having easier switching condition. We thus conclude that the high quality of our sample, indicated by the large and steep transition (low dispersion of critical temperature among the domains), is a determining factor in the observed behaviour.
	
	In summary, we analysed the resistive switching effect in \ce{VO2} microbridges under strong current biasing conditions. The observed sharp resistance drops are in contrast with a simple model based on a homogeneous-medium approximation and the negative differential resistance of the \ce{VO2} layer. We interpreted the observed switching phenomena considering the local redistribution of current density due to the pinch effect of randomly created metallic grains within the microbridge. The comparison of the characteristics of the resistive jumps in different heat dissipation conditions supports the thermal origin and the intrinsic randomness of the observed behaviour. Our results suggest that the reduction of out-of-plane dissipation in free-standing devices can be an efficient tool to suppress abrupt jumps of the electrical resistance with increasing current, facilitating the gradual control of \ce{VO2}-based devices.  In addition, the intrinsic randomness in the formation of conducting paths evidences that particular care should be taken when modelling the switching behaviour of thin film \ce{VO2}-based devices.	
	\\
	\\
	We acknowledge the financial support by a Grant-in Aid for Challenging Exploratory Research (No. 26600074), a Grant-in-Aid for Scientific Research B (No. 25286058) from the Japan Society for Promotion of Science (JSPS), and FIRB RBAP115AYN ``Oxides at the nanoscale: multifunctionality and applications'', Executive programme of cooperation between Italy and Japan by the Italian Ministry of Foreign Affairs, PRIN 2010NR4MXA ``OXIDE''. N.M. also thanks  G. Mattoni and D. J. Groenendijk for useful discussions.

	\bibliographystyle{apsrev4-1}
	\bibliography{library.bib}	

\end{document}